# Molecular Dynamics Simulation of Bubble Nucleation in Hydrophilic Nanochannels by Surface Heating

*Manish Gupta & Shalabh C. Maroo\**
Department of Mechanical Engineering and Aerospace Engineering, Syracuse University, Syracuse, NY 13244; Corresponding author email: scmaroo@syr.edu

**Abstract:** Bubble nucleation in liquid confined in nanochannel is studied using molecular dynamics simulations and compared against nucleation in the liquid over smooth (i.e. without confinement). Nucleation is achieved by heating part of a surface to high temperatures using a surface-to-liquid heating algorithm implemented in LAMMPS. The surface hydrophilicity of nanochannels is increased to understand its effect on nucleation behavior. Liquid structuring is found to play a significant role in altering thermodynamic properties of density and pressure in the nanochannels, which in turn changes the enthalpy of vaporization. Increased surface hydrophilicity in nanochannels results in the delay of bubble formation as more energy is required for nucleation. Thus, bubble nucleation in hydrophilic nanochannels can dissipate higher heat fluxes and can potentially be used towards the thermal management of hot spots in power electronics.

**Main Text**

Heat dissipation from a hot spot is the primary challenge in thermal management [1, 2] of power electronics and high precision instruments in order to maintain a safe and desired surface temperature. Recent literature shows a significant amount of work is being done to dissipate high heat flux from such electronics, and solutions include creating micro/nano-patterns on the surface [3-7] coupled with liquid-vapor phase change as it is one of the most efficient processes in heat transfer [8]. A few studies have suggested nanochannels can significantly enhance heat flux dissipation from surfaces [3, 9]. Nucleation in confined liquid can be different from that without confined liquid due to changes in thermodynamic properties and structuring of liquid atoms near the surface [10]. An experimental study by Lin [11] investigated the formation of a bubble in microchannels and concluded that homogenous bubbles are more likely to form in microchannels. Even though the classical theory predicts the formation of heterogeneous nucleation due to the presence of surfaces, a highly smooth surface may not be able to initiate bubble nucleation [11, 12]. At nano-length scales, liquid structuring at the surface can affect nucleation behavior while keeping the surface wet at higher temperatures. Studies have shown that heat transfer is higher when surface is wetted [13, 14] due to evaporation of liquid on heated spots. When dry region forms, the local heat flux suddenly drops [15, 16] and can cause thermal failure of the surface. Yu et al. [17] used molecular simulations to show that bubble nucleation temperature increases when the size of the channel is reduced, as the effect of increasing confinement is similar to having surface-liquid interaction dominate the interatomic interaction resulting in a dense structured layer. Hydrophilic surfaces can increase the wettability of surfaces [14] and delay the formation of a dry region by keeping the liquid layers on a heated surface through surface-driven passive liquid flow [18]. Novak et al. [14] investigated homogenous and heterogeneous nucleation at the surface using molecular simulations and showed that nucleation occurs at lower temperatures when surface-liquid interaction is weak. This work focuses on studying heterogeneous bubble formation inside nanochannels, using molecular dynamics (MD) simulation, as a potential solution towards hot spot thermal management. A fixed geometry nanochannel is used and confinement effects are varied by changing the strength of surface-liquid interaction. We identify the variation in thermodynamic properties and heat transfer characteristics of nucleation in nanochannels compared to that in non-confined liquid over smooth surface. As heat from the surface is removed by evaporation, change in enthalpy of fluid before nucleation and after nucleation is also estimated as it serves as an important parameter [8] to predict heat dissipation characteristics.

First, bubble nucleation on surface without any nanochannel confinement is studied as the baseline case. The simulation domain consists of 30 nm thick liquid argon over 25 nm x 5 nm platinum surface (Fig. 1a). A 20 nm region of argon vapor atoms along with 2 nm of liquid argon on an identical platinum surface are placed over liquid argon to help stabilize bulk properties during the simulation run. The simulation was initially run for 5 ns at 110K using Nosé–Hoover thermostat to attain equilibrium with the following parameters [18]: $\varepsilon_{Ar-Ar}$ = 1.0179672 kJ/mol, $\sigma_{Ar-Ar}$ = 0.34 nm and $\varepsilon_{Ar-Pt}$ = 0.5385 kJ/mol, $\sigma_{Ar-Pt}$ = 0.3085 nm. The time step was 5 fs and all simulations were run in LAMMPS software [19]. After 5 ns, the thermostat was turned off and all the platinum surfaces were kept at a constant temperature of 110K for 1 ns using an in-house implemented surface-to-fluid heat transfer algorithm [20, 21].

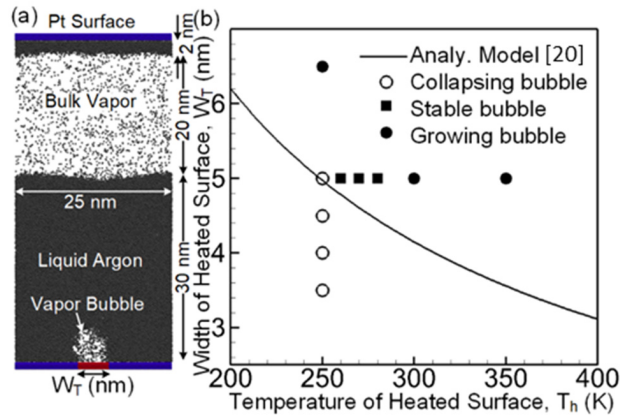

Figure 1: Bubble nucleation in liquid argon over smooth surface showing (a) simulation domain, and (b) bubble growth behavior for different heating widths and surface temperatures.

A part of the platinum surface ($W_T$) is heated to a higher temperature to initiate nucleation in the liquid; the heat transfer algorithm enables heating/cooling different parts of the same surface to different temperatures. A total of 10 cases are run for varying widths from 3.5 nm to 6.5 nm and heated surface temperatures varying from 250 K to 350 K. For a lower temperature of the heated surface ($T_h$) and smaller heating width, the bubble is found to initially grow but it eventually collapses. An analytical model [20] is used to predict such a heterogeneous nucleation behavior and for comparing against our simulation results. The thermodynamic properties of model and detailed validation can be found on our previous work [20].

As seen in Fig. 1b, the heating width and surface temperature conditions below the analytical model result in the bubble collapsing over time. The bubble remains stable for conditions around the model prediction while it grows for conditions above the analytical model line. This phenomenon shows the existence of critical radius of nucleation as predicted by the classical theory of nucleation [8]. The obtained results show that the simulation results are in good agreement with the analytical model prediction.

Next, to study bubble nucleation inside a nanochannel with varying hydrophilicity, the strength of surface-liquid interaction is increased through the parameter ε and captured using a non-dimensional parameter $\varepsilon^* = \varepsilon/\varepsilon_{Ar-Pt}$ where $\varepsilon_{Ar-Pt}$ = 0.5385 kJ/mol as defined earlier. However, changing surface-liquid interaction also affects the critical distance $R_{cr}$ used to achieve the desired surface temperature in the surface-to-fluid heat transfer algorithm as $R_{cr}$ depends on the strength of surface-liquid interaction [21] (See supporting information). Thus, to determine the critical distance required to heat the surfaces of different hydrophilicities, a MD simulation domain shown in the inset of Fig. 2 is used where liquid argon is sandwiched between two platinum surfaces. The temperature of the lower wall and upper wall are maintained by the heat transfer algorithm. Initially, both the wall temperatures are at 90 K for equilibration following which the upper wall temperature is increased to 140 K while the lower wall temperature is kept at 90 K. Four cases are run for $\varepsilon^*$ equaling 1, 2, 4, and 6, and variation of the average temperature of liquid argon is plotted against time. As the domain resembles a 1D heat conduction problem, the temporal variation of average temperature of liquid argon for all cases are compared with the analytical solution of the 1D heat equation (Fig. 2) and are found to be in good agreement. The corresponding $R_{cr}$ values used for these four cases are plotted in the inset of Fig. 2 and the critical distance is found to decrease with increasing hydrophilicity. Thus, these values of $R_{cr}$ can be used to simulate heat transfer from such surface with varying hydrophilicity (i.e. ε*), and are used next to study bubble nucleation in nanochannels.

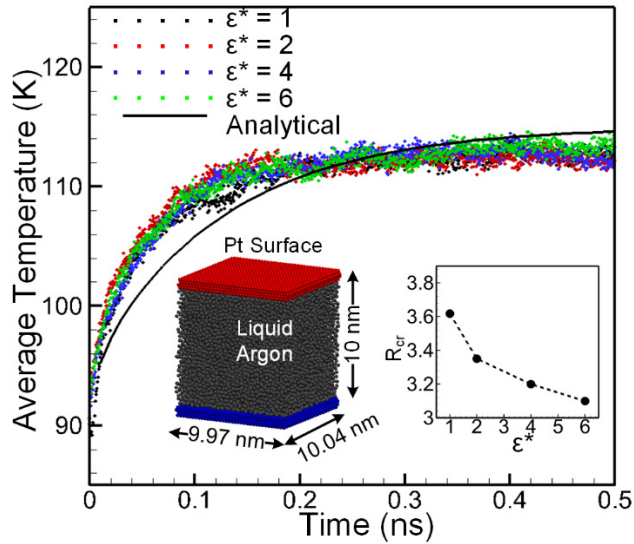

Figure 2: Average bulk liquid temperature variation over time using implemented surface-to-fluid heat transfer algorithm and compared with 1D heat condition equation to determine critical distance $R_{cr}$ for increased surface-liquid interaction $\varepsilon^*$.

In order to study bubble nucleation inside a nanochannel, a new platinum wall is placed at a distance of 5 nm above the lower wall in the simulation domain of Fig. 1a. The modified domain is shown in Fig. 3a. The other domain parameters are the same as the previous simulation of Fig. 1a. Four simulation cases are run with increasing hydrophilicity of the surface based on $\varepsilon^*$ values of 1, 2, 4, and 6. Each simulation is equilibrated for 2 ns using Nose-Hoover thermostat at 110 K followed by turning off the thermostat and maintaining surface temperatures at 110 K for another 1 ns using the surface-to-fluid heat transfer algorithm. In the equilibrium state, the density and pressure of liquid argon in the nanochannel are plotted in Fig. 3b; the same properties of liquid argon without the nanochannel are also included in the same plot (estimated from Fig. 1 simulations). Compared to the no-channel case (marked as 'NC'), the liquid pressure and density slightly increase when the channel is present (marked as '5 nm C') due to liquid structuring. This effect is enhanced within the channel with increasing liquid-surface interaction $\varepsilon^*$. Figure 3c shows the absolute enthalpy values of liquid argon with confinement during the equilibrium period. As the liquid is more structured for higher values of $\varepsilon^*$, the magnitude of enthalpy increases. Thus, thermodynamic properties in confinement obtained by changing hydrophilicity shows that more hydrophilic surfaces mimic the behavior of increasing confinement, i.e. smaller channel heights where liquid structuring further dominates the overall properties.

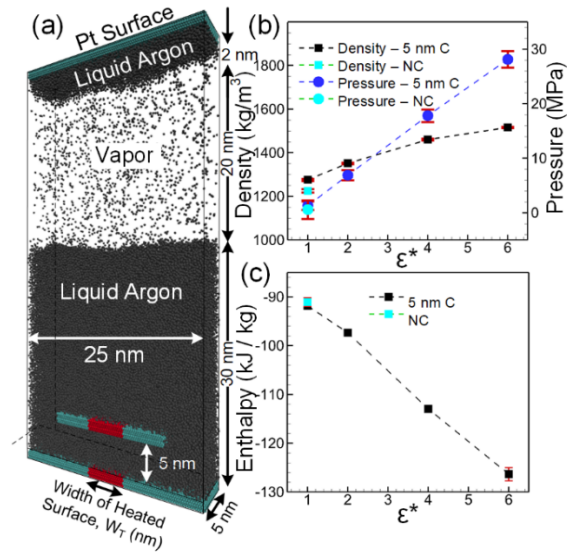

Figure 3: Equilibrium simulations with 5 nm-height nanochannel showing (a) simulation domain, (b) density and pressure variation with $\varepsilon^*$, and (c) absolute enthalpy of liquid argon variation with $\varepsilon^*$.

After equilibration, the 5 nm width $W_T$ of the lower and upper walls of the nanochannel, shown in Fig. 3a, is increased to a higher temperature of 350 K to initiate bubble formation in the nanochannel. The rest of the surfaces in the domain are maintained at the initial equilibrium temperature of 110 K. A bubble nucleates and forms at the heated part of the walls. The evolution of a vapor bubble is shown in Fig. 4 for the no-channel case (with $\varepsilon^*$ = 1, Fig. 4-a) as well as nanochannel cases with $\varepsilon^*$ = 1 (Fig. 4-b). The density distribution is calculated by dividing the domain into 0.25 nm x 0.25 nm rectangular bins and averaging the fluid properties over every 50 ps. The bubble grows with time for the no-channel case and the dry region forms over the heated area. At 0.1 ns (Fig. 4-a1), the bubble size is smaller but distinct, and grows to a much bigger bubble at 1.0 ns (Fig. 4-a3); the corresponding density contours confirm this observation. However, for the same surface-liquid interaction ($\varepsilon^*$ = 1) in the nanochannel, the bubble does not distinctly form at 0.1 ns (Figs. 4-b1 and 4-b2). Thus, this clearly shows that nucleation is difficult to achieve in nanochannel because of higher density and pressure (Fig. 3b) in the confined liquid. It can also be seen that at 1 ns (Figs. 4-b3 and 4-b4), vapor bubble forms but with a density higher than the no-channel case (Fig. 4-a4 vs. Fig. 4-b4). Thus, such effects are expected to become more prominent as surface hydrophilicity is increased.

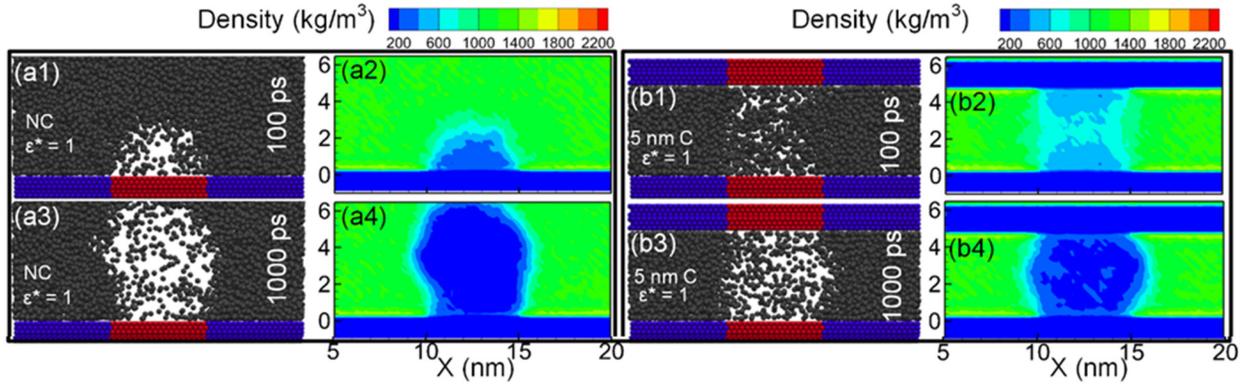

Figure 4: Evolution of vapor bubble using molecular visualization and density contours at initial bubble growing phase (0.1 ns) and final steady-state (1.0 ns) for (a) Not confined liquid with no-channel for $\varepsilon^* = 1$, and (b) 5 nm-height nanochannel for $\varepsilon^* = 1$,

Figure 5 shows the growth of the bubble in the nanochannel for $\varepsilon^* = 6$. At 0.1 ns, the liquid argon heats up but does lead to bubble formation (Figs. 5-c1 and 5-c2). With time, the bubble grows between the heated walls resulting in an elongated oval shape (Figs. 5-c3 and 5-c4). Interestingly, a liquid monolayer exists on the surface unlike the other cases of Figs. 4a and 4b; the presence of the monolayer is due to the very strong surface-liquid interaction.

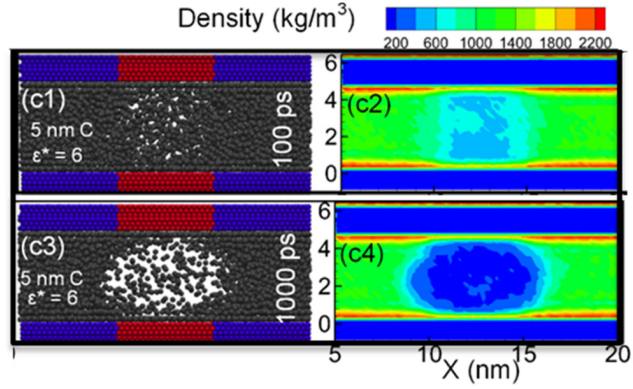

Figure 5: Evolution of vapor bubble using molecular visualization and density contours at initial bubble growing phase (0.1 ns) and final steady-state (1.0 ns) for 5 nm-height nanochannel for $\varepsilon^* = 6$ (Hydrophilic surface).

At steady state, the phase change continues with evaporation happening over heated parts of the walls and condensation at the relatively cooler ends of the bubble (See supporting document for more information). It is expected that heat transfer would be higher in the channels where the surface is more hydrophilic. The qualitative analysis of heat dissipation is done by estimating the change in enthalpy of vaporization as follows:

$$\Delta h = \frac{\sum(N_v * (h_v - h_0))}{\sum(N_v)} \qquad (1)$$

where, $N_v$, is the number of vapor atoms, $h_v$ is enthalpy of vapor atoms, and $h_0$ is the initial enthalpy of liquid argon in the channel before heating was started. Further, enthalpy $H$ is the sum of internal energy $U$ and work required to achieve its pressure $P$ and volume $V$ as follows [8]:

$$H = U + PV \qquad (2)$$

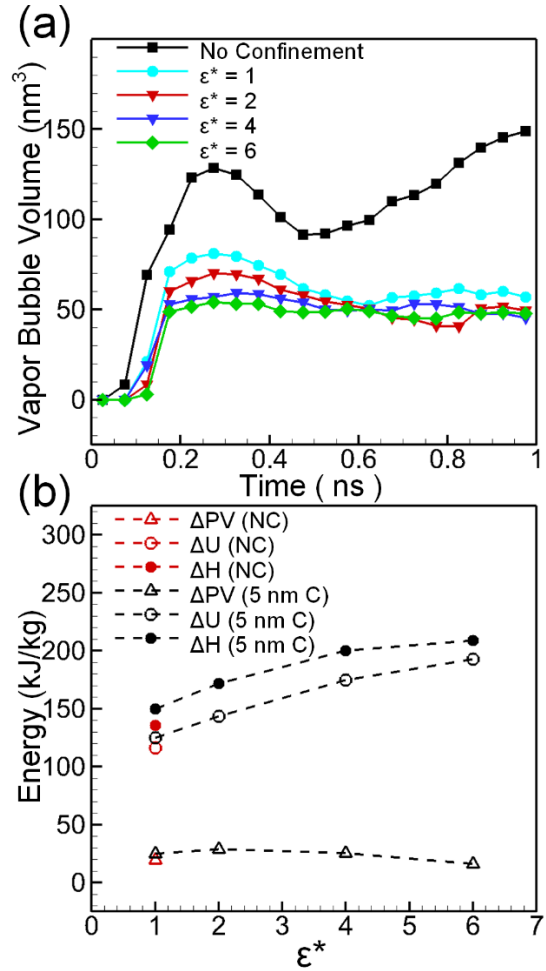

Figure 6: Bubble growth and energy characteristics of bubble nucleation in nanochannel showing (a) vapor bubble volume variation with time, and (b) change in enthalpy, internal energy and work required for bubble nucleation with confinement and hydrophilicity of the surface.

The change in enthalpy represents the heat gained in the fluid from the heated walls, and a higher value of $\Delta h$ implies higher heat dissipation from the walls. The enthalpy of vapor atoms and initial liquid atoms were calculated by a statistical ensemble of atomic properties in 0.25 nm x 0.25 nm rectangular bins. The variation of vapor bubble volume with time is shown in Fig. 6a. In the case of no-channel, a larger bubble forms early due to a sudden increase of surface temperature and lack of confinement. The unsteady large bubble leads to immediate condensation of vapor at the bubble top resulting in a slight decrease in bubble size. Once a quasi-steady state is reached, the bubble continuously grows over time as the fluid gains energy from the heated surface. In the case of nanochannel, once the bubble reaches a steady-state, the bubble size remains almost similar for all cases as the bubble is restricted inside the channel and condensation occurs at the liquid-vapor interface away from the surface. The change in thermodynamic energies of the system is shown in Fig. 6b, where Δh is calculated by equation *(1)* with the help of equation *(2)*. The change in U and PV are statistically obtained from MD simulations and are also plotted separately to show the importance of each term during nucleation. The change in expansion energy term (PV) remains almost the same for all cases, slightly decreasing for higher ε* cases as the vapor bubble volume V decreases (Fig. 6a) enough to counteract the increase in pressure P. The internal energy part of enthalpy change is greater for the nanochannel than nucleation in without confinement for the same surface-liquid strength (ε*=1), and it increases almost linearly with increase in $\varepsilon^*$. Similarly, the change in enthalpy of the system follows the internal energy trend. The increase in enthalpy of vaporization for such confined nanochannels coupled with hydrophilic surfaces shows that heat dissipation can be enhanced by over 30%, and such a combination can potentially be utilized towards the thermal management of electronics.

To summarize, molecular dynamics simulations are used to study vapor bubble nucleation and heat transfer in 5 nm height nanochannels with increasing surface hydrophilicity while comparing against bubble nucleation in without confined liquid. Nanochannel confinement and increase in strength of surface-liquid interaction results in change in liquid density and pressure as more liquid structuring occurs next to the surface. Nanochannel confinement is found to require more energy to convert liquid into vapor atoms and is captured through an increase in enthalpy of vaporization. The effect is more visible when the relative strength of surface-liquid interaction is increased based on multiples of 2, 4 and 6. Hence, nucleation in nanochannels can result in higher energy dissipation from the surface with increasing confinement effects and can be used to advance thermal management solutions.

**Acknowledgement:** This material is based upon work supported by, or in part by, the Office of Naval Research under contract/grant no. N000141812357.

**Data Availability:** The data that support the findings of this study are available from the corresponding author upon reasonable request.